# Josephson current via spin and orbital states of a tunable double quantum dot

Rousan Debbarma[1], Markus Aspegren[1], Florinda Viñas Boström[1,3], Sebastian Lehmann[1], Kimberly Dick[2], Claes Thelander[1,*]

[1]*Division of Solid State Physics and NanoLund, Lund University, Box 118, S-221 00 Lund, Sweden*

[2]*Center for Analysis and Synthesis, Lund University, Box 124, S-221 00 Lund, Sweden*

[3]*Institut für Mathematische Physik, Technische Universität Braunschweig, D-38106 Braunschweig, Germany*

*Abstract.* Supercurrent transport is experimentally studied in a Josephson junction hosting a double quantum dot (DQD) with tunable symmetries. The QDs are parallel-coupled to two superconducting contacts and can be tuned between strong inter-dot hybridization and a ring geometry where hybridization is suppressed. In both cases, we observe supercurrents when the two interacting orbitals are either empty or filled with spins, or a combination. However, when each QD hosts an unpaired spin, the supercurrent depends on the spin ground state. It is strongly suppressed for the ring geometry with a spin-triplet ground state at zero external magnetic field. By increasing the QD hybridization, we find that a supercurrent appears when the ground state changes to spin-singlet. In general, supercurrents are suppressed in cases of spin doublet ground state, but an exception occurs at orbital degeneracy when the system hosts one additional spin, as opposed to three, pointing to a broken particle-hole symmetry.

*Introduction.* Semiconductor quantum dots (QDs) offer unique possibilities to study the interactions of various localized spin and orbital electron states with the macroscopic properties of a superconductor. Confining a single-electron spin to a QD provides a controlled magnetic impurity, which can couple to quasiparticles in the superconductor to give rise to various states within the superconducting gap [1–4]. It was recently shown that such sub-gap states can provide a basis for qubit implementation [5]. In systems with double QDs (DQDs), electron spins can be delocalized or confined to each of the QDs, and the interactions between the QDs and their surrounding can be tuned over a wide range. Andreev molecular levels have been experimentally studied in serial DQDs formed by local gating of semiconductor nanowires [6,7], and screening of DQD spin-doublet and singlet states has been demonstrated by tuning the tunnel couplings to a superconductor [8–10].

In Josephson junctions involving QDs, dissipationless supercurrents can flow not only at QD resonances but also in Coulomb blockade as a result of higher-order transport processes [11]. When the ground state of the QD is a doublet, the Josephson relation can acquire a $\pi$ phase shift [11,12] which can be detected by embedding the QD in a SQUID. Recently, the critical current of a serial DQD Josephson junction was explored, where its dependence on occupancy and level detuning was mapped [13]. In a similar system, a supercurrent suppression was reported when the two-electron spin ground state was changed from singlet to triplet by an external magnetic field [14]. Further, in tunneling spectroscopy of a serial DQD, the triplet ground state was suggested to inhibit Andreev reflection to the superconducting contact [15].

Systems in which superconductors instead interface parallel DQDs have been of particular interest for understanding non-local tunneling processes of Cooper pairs [16]. This configuration has more degrees of freedom and does not require a finite inter-dot coupling for transport. Given the range of tunable parameters and possible interference effects, the geometry has been a subject of numerous theoretical studies [16–18]. Experimental works have focused on transport in parallel DQDs with no, or very small, inter-dot tunnel coupling, such as on Cooper pair splitting [19], as well as on transport via Andreev bound states in QDs located in closely spaced nanowires [20,21].

In this work, we explore supercurrent transport through a parallel-coupled DQD, where the two-electron spin ground state can be controlled electrostatically. We thereby avoid magnetic-field-induced reduction of the Josephson effect as well as Zeeman splitting of spin-triplet states. By coupling the QDs in two separate points, we can tune the effective interdot-hybridization and the exchange interaction that determine the spin-singlet and triplet energies. A strong suppression of the critical current is found when the DQD ground state changes from spin-singlet to triplet. In general, Cooper pair transport primarily seems to take place through empty and filled orbitals. However, in agreement with theoretical predictions [18], we find an exception at orbital degeneracy where the electron-hole symmetry is broken in the supercurrent transport involving odd electron states. Finally, we observe clear subgap levels that can be correlated to spin states of the DQD, which indicate a connection between the energies of subgap states and the suppression of critical currents in the two-electron regime.

*Sample fabrication and quantum dot formation.* Electron transport was studied in an InAs(Sb) nanowire with approximately 80 nm InAs core and 4 nm $InAs_xSb_{1-x}$ shell as illustrated in Fig. 1 [22]. A QD was formed during the epitaxial nanowire growth by introducing a pair of closely spaced wurtzite (WZ) segments, acting as tunnel barriers, along the length of otherwise zinc blende (ZB) nanowire segments [23]. In the electrical measurements, such a thin, disk-shaped QD can be further split into coupled QDs, situated near the nanowire surface [24]. The role of the InAsSb surface layer is to enhance the electron concentration and reduce series resistances. A Josephson junction was fabricated by depositing a Ti/Al (5/90 nm) film after defining electrode patterns via electron beam lithography. DC transport measurements (see Fig. S3 for circuit diagram) were performed at 40 mK base temperature in a dilution refrigerator equipped with a vector magnet.

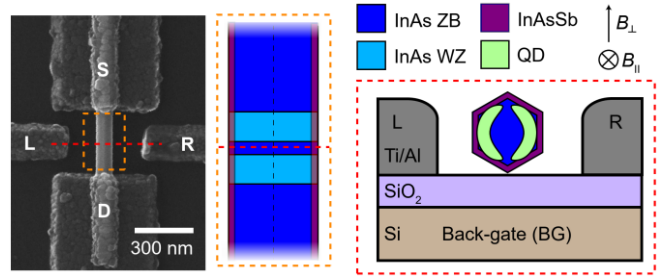

FIG. 1. Scanning electron micrograph of a typical device in the study. An InAs(Sb) quantum dot embedded inside a nanowire is connected to source (S) and drain (D) contacts of Ti/Al which become superconducting at temperatures below 1 K. The QD confinement is provided by tunnel barriers of wurtzite (WZ) InAs. Two additional electrodes (L, R) act as side gates, which together with a back-gate (BG) control the potential such that the QD can be split into coupled QDs.

In this study, we induced two QDs in a nanowire by tuning voltages applied to a back-gate and two side-gate electrodes. We note that the QDs, here connected in parallel with source and drain, can couple to each other at two separate points along the nanowire circumference (Fig. 1). We have previously shown that ring-like DQDs can form in situations where the two tunnel couplings are of similar magnitude and where the overlap integrals have different sign [24]. We note that the latter condition is fulfilled when pairs of even and odd orbitals hybridize. In such a situation, the hybridization energy of the interacting orbitals cancel, and despite large individual tunnel couplings, the orbitals become nearly degenerate, split mainly by spin-orbit interaction. Such a ring-like symmetry provides a very large orbital contribution to the energy of a state in a magnetic field, and also results in a spin-triplet ground state (GS) when each QD hosts an unpaired spin in the highest unfilled orbital, in agreement with Hund's rule [25]. However, by upsetting the tunnel coupling balance, a conventional DQD can be formed, where the orbital contribution to the $g$-factor is quenched, and where the conventional spin-singlet GS of a tunnel-coupled DQD is regained.

*Weak orbital hybridization.* We start by investigating how the critical current of a ring-like DQD Josephson junction depends on charge state, and later we modify the same orbitals into a conventional tunnel-coupled DQD configuration.

Conductance, $G$, plotted as a function of voltages applied to the two side-gates ($V_L$ and $V_R$) is shown in Fig. 2(a) for a fixed back-gate ($V_{BG}$). The figure shows two orbitals that come close in energy and weakly interact as seen by the shape of the corners of the honeycomb [26]. From overview measurements [27], we estimate that the DQD here has approximately 30 electrons in filled orbitals. Differential conductance, plotted as function of the applied source-drain voltage ($V_{bias}$) passing through the four triple points (red line in Fig. 2(a)) of the orbital crossing, is shown in Figs. 2(b,c). In Fig. 2(b), a $B$-field out of the substrate plane ($B_\perp$) has been applied to quench the superconductivity in the leads.

The measurements show Coulomb diamonds inside of which the charge state is constant, going from a (0,0)e to a (2,2)e DQD charge configuration, where we have omitted electrons in filled orbitals at lower energy. Transport inside Coulomb diamonds is possible through co-tunneling processes where the charge state does not change. Here, a transition to a darker color signifies that transport through an excited state is possible.

In the (1)e and (3)e diamonds [Fig. 2(b)], the low-energy excited states are explained by the two nearly degenerate orbitals that spin-split in the weak $B$-field. Transport in the (1,1)e charge state looks very similar, although here the excited state spectrum is explained by a triplet GS, which spin splits, and a singlet excited state at higher energy. Evidence for this atypical ordering of the two-electron states is presented later. From the Zeeman-induced spin-splitting, we extract a spin $g$-factor component, $g_s \approx 9$. At $B = 0$ [Fig. 2(c)] we note the presence of a superconducting gap ($\Delta \approx 190$ μV), which results in peaks in the d$I$/d$V_{bias}$ at both $\Delta$ and 2$\Delta$, as well as a strong suppression (white) for $|V_{bias}| < \Delta$ in Coulomb blockade.

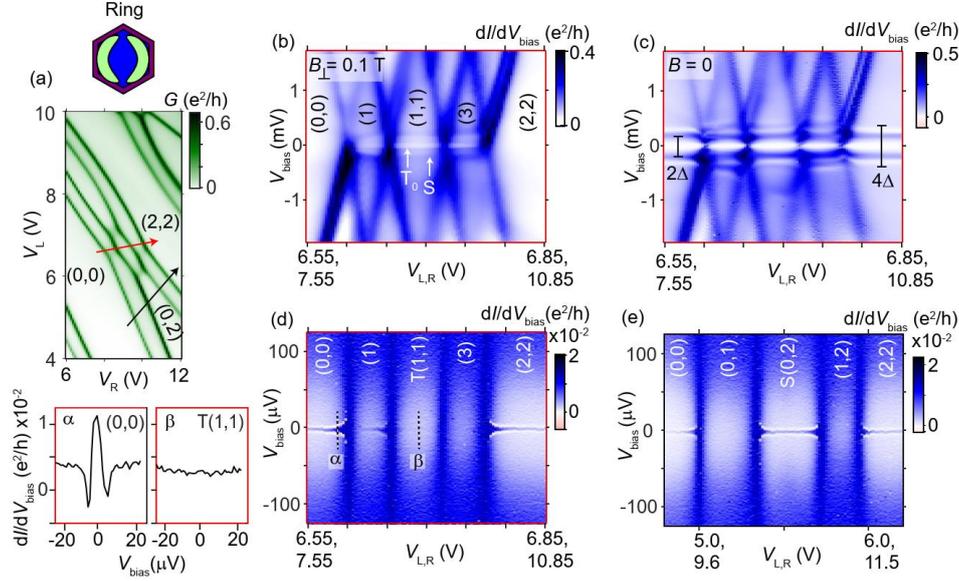

FIG. 2. (a) Conductance vs. $V_L$ and $V_R$ for a fixed $V_{BG}$ = -3.3 V in a regime of symmetric potentials where the DQD has ring-like properties ($V_{bias}$ = 0.33 mV). The DQD here has approximately 30 electrons in filled orbitals. (b) Differential conductance as function of $V_{bias}$ (stability diagram) along the red gate vector ($V_{L,R}$) crossing the four triple-points in the DQD honeycomb. Co-tunneling transport processes involving excited DQD states give rise to horizontal features within the Coulomb diamonds. A magnetic field ($B_\perp$) is applied to quench the superconductivity in the leads and with a direction perpendicular ($\perp$) to the NW axis where no flux threads the ring. (c) Corresponding plot at $B$ = 0 where the superconducting gap is visible. (d) Supercurrent measurement ($R_S$ = 1.7 MΩ) across the meeting point of the orbitals (red vector), where we note a suppression of the supercurrent when each orbital has an unpaired spin. The two line-cuts at positions α and β show the relative effect of this suppression in (1,1) compared to (0,0). At α the critical current is ~90 pA, which increases towards the DQD charge degeneracy. (e) Corresponding measurement along the black gate-vector outside the honeycomb where each QD (orbital) is filled sequentially.

*Supercurrent transport under weak orbital hybridization.* By adding a series resistance ($R_S$ = 1.7 MΩ) and reducing the voltage $V_{bias}$ applied over the setup (including $R_S$), we next explore supercurrent transport through the DQD. $R_s$ acts to limit the current for a given $V_{bias}$ and at the same time reduces the voltage drop over the device, which we take into account when extracting energies. Figure 2(e) shows differential conductance along the black line in Fig. 2(a), outside of the DQD honeycomb, where one QD orbital at a time is filled with electrons. For some charge configurations, we observe a strong signal around zero bias, surrounded by low or negative differential conductance features (see Fig. 2d). This feature, which is now resolvable due to $R_S$, indicates supercurrent [17,25] with a magnitude that can be extracted by fitting the measurement to an RCSJ model [27]. However, in this work we focus primarily on the absence or presence of such a feature in relation to the spin and orbital configuration of the DQD. Figure 2(e) shows the presence of a supercurrent in the even-electron regimes where each QD orbital is either empty of, or filled with, electrons. This would be comparable to a 0-junction for a single QD embedded in a Josephson junction. However, the supercurrent is strongly suppressed in the odd-charge configurations. Such a suppression is commonly observed experimentally in a π-junction [11–13,28]. Theoretical calculations have shown that odd electron states form π-junctions and the presence of both spin order preserving and flipping transport processes lead to Josephson current suppression [29–31].

By instead crossing through the four triple points [red line, Fig. 2(d)], the situation looks different. A supercurrent is still present in the (0,0)e and (2,2)e regimes where both QD orbitals are either empty or filled, but we note a strong suppression in the (1,1)e configuration with a spin-triplet GS, and oppositely the emergence of a weak supercurrent in (1)e, but not in (3)e.

Assuming that local transport of Cooper pairs dominate, such that two electrons pass via the same QD, the triplet-related suppression may be explained by physics similar to the spin 1/2 suppression in a single QD [29–31]. However, if we also consider non-local transport, i.e. a splitting of the Cooper-pairs, a large number of higher-order tunneling processes become possible. Probst et al. [16] studied non-local transport in a parallel DQD system with indirect tunneling between the QDs, and predicted that the spin-triplet state supports a supercurrent, although weaker in magnitude than the singlet. The suppression was explained by Pauli spin blockade that reduces the number of allowed transport paths for the spin-triplet GS.

Concerning the appearance of a supercurrent in (1)e but not (3)e, we find that the same asymmetry appear in other orbital crossings, with examples given in Fig. S1(e) and S2(e) in the Supplemental Material [27]. In the case of a DQD with identical QD-superconductor couplings, Droste et al. [18] predicted a broken particle-hole symmetry of the supercurrent at vanishing level detuning. Due to interference effects, the anti-bonding orbital becomes decoupled from the superconductor, such that only the bonding orbital carries a supercurrent. Scherübl et al. [32] recently predicted a broken electron-hole symmetry in the level structure of sub-gap states of a similar system. They found that an asymmetry appeared in the special case where the QDs couple both directly and indirectly through the superconductor, and where non-local tunneling is also possible.

*Strong orbital hybridization.* Next, by changing $V_{BG}$ with +1 V, and compensating with $V_{L,R}$, we modify the shapes of the orbitals for the same DQD charge configuration. Now, the system behaves as a typical DQD in the limit of strong inter-dot coupling, as evidenced by the avoided orbital crossing in Fig. 3(a). A stability diagram [$B = 0$, Fig. 3(b)] along the line in Fig. 3(a) shows a doublet excited state (antibonding) in the (1)e regime, and a triplet excited state in (1,1)e. A corresponding supercurrent measurement is shown in Fig. 3(c), where we find a supercurrent in (1,1)e where the GS now is spin-singlet. In a simplistic non-interacting picture, this corresponds to having the bonding-orbital filled with two electrons, where therefore a 0-junction behavior may be expected.

Figures 3(d,e) represent demonstrations of the electrostatic tuning of the (1,1) GS from the ring-like triplet to a conventional DQD singlet, i.e. going from the center of the honeycomb in Fig. 2(a) and crossing through the corresponding center of Fig. 3(a). The gate trajectory was chosen as to remain within the (1,1) configuration, and determined by first obtaining conductance plots similar to Fig. 2a and 3a for various $V_{BG}$. In Fig. 3(d), a $B_\perp = 100$ mT is applied to quench the superconductivity during this operation. For the $T_+$ GS (left side) we note transport through two excited states, $T_0$ and S, ($T_-$ is not accessible with exchange of only one spin). However, for the singlet GS (right side), transport involving all three triplet excited states is possible, resulting in an additional resonance (see Fig. S5). This also explains the excited state spectrum inside the (1,1) Coulomb diamond in Fig. 2a. Changing to a supercurrent measurement configuration [Fig. 3(e), $B = 0$], we find that a supercurrent appears when the ground-state changes from T to S.

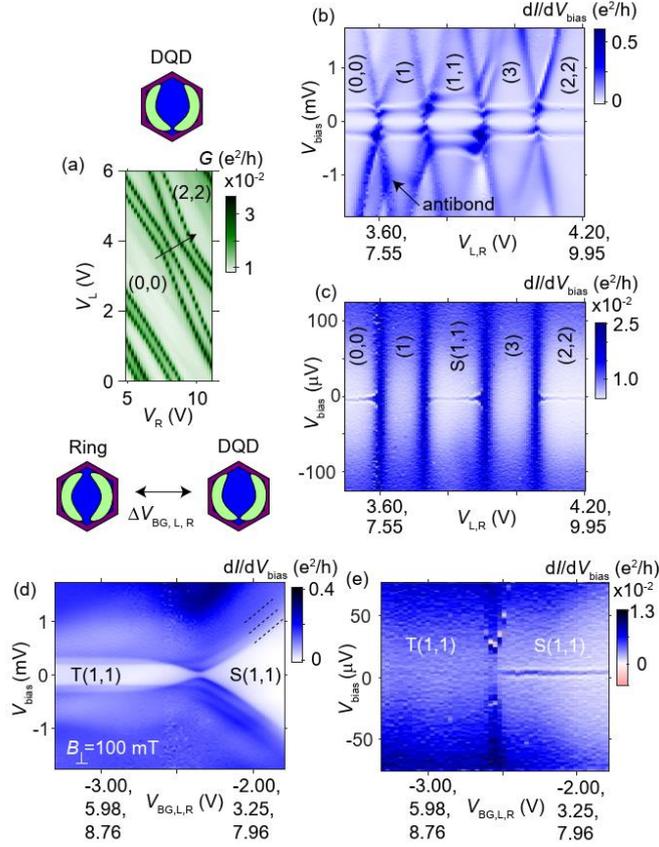

FIG. 3. Conductance vs. $V_{L,R}$ for a fixed $V_{BG}$ = -2.3 V in a regime where the QDs strongly couple in one point ($V_{bias}$ = 0.17 mV). (b) Stability diagram along the vector in panel a where we find a one-electron excited state (antibonding) at high energy that confirms a strong orbital hybridization. (c) Corresponding supercurrent measurement where we note supercurrent transport near zero bias involving the S(1,1) ground state of the hybridized DQD. (d) Differential conductance vs. $V_{bias}$ inside the (1,1) honeycomb during electrostatic tuning ($V_{BG, L, R}$) of the two-electron GS. Left side corresponds to the ring-like case in Fig. 2 with a T(1,1) GS, and right to a conventional DQD with S(1,1) GS. (e) Corresponding supercurrent measurement where a supercurrent appears as the GS changes from T to S.

*B-field tuning of the (1,1) spin ground state.* Next we go back to the DQD ring configuration, where we take advantage of the strong *B*-field dependence of the S(1,1) state for fields applied parallel to the nanowire long-axis ($B_∥$). Here, a spin triplet-singlet GS change can be induced before quenching the superconductivity in the leads. We point out that the orbital properties of this singlet is very different from the corresponding singlet in the conventional DQD.

Figure 4(a) shows the ground state evolution as function of $B_∥$ recorded along the red line in Fig. 2(a) that passes through the four triple points. The rapid evolution of the states confirms the ring-like nature of these orbitals [22], where the orbital contribution to the Zeeman splitting corresponds to a component $g_{orb}$ ≈ 170. Such large and anisotropic *g*-factors have been observed previously when forming rings of various symmetry within similar nanowire QDs [22]. The strong dependence on $B_\parallel$ of the lowest energy singlet S(1,1) is shown in Figs. 4(b,c), where in Fig. 4(b) also a $B_\perp$= 100 mT was applied to quench the superconductivity. A ground state transition occurs at $B_\parallel$ = 25 mT where the singlet becomes GS. The corresponding supercurrent measurement is shown in Fig. 4(d), although it contains no indications of a supercurrent despite the GS change. At this $B$-field, the supercurrent is not fully quenched for even-electron configurations elsewhere in the honeycomb [as shown later in Fig. 5(h)], which may point to a suppression of supercurrent through this singlet.

A supercurrent measurement as function of $B_\parallel$ in the (1,1)e for another orbital crossing of the same device with a ring-like behavior is shown in Figure 4(e). In this case the DQD has a less stable triplet GS ($E_S$ - $E_T$ = 80 µeV instead of 160 µeV) in the normal state. Here we note a weak supercurrent regardless of spin GS in the normal state (see Supplemental Material [27]). Instead of decaying with $B$-field, the supercurrent signal stays nearly constant, or potentially increases, as S(1,1) becomes lower in energy than T(1,1). As reference, we compare with a measurement in the S(0,2) regime of the same orbital crossing [Fig. 4(f)], where the supercurrent signal is correspondingly stronger, and where we note the expected decay of the signal with applied $B$-field.

Similar to recent findings on screening of spin-singlets in serial DQDs [8–10], the spin-triplet should also become screened by quasiparticles in the superconductor for sufficiently strong coupling to the leads. We find that an excited state close to zero bias [Fig. 4(e)], possibly associated with such screening, is a reoccurring feature in the (1,1) regime of other ring-like orbital crossings where the supercurrent is not fully suppressed (See Supplemental Material [27]).

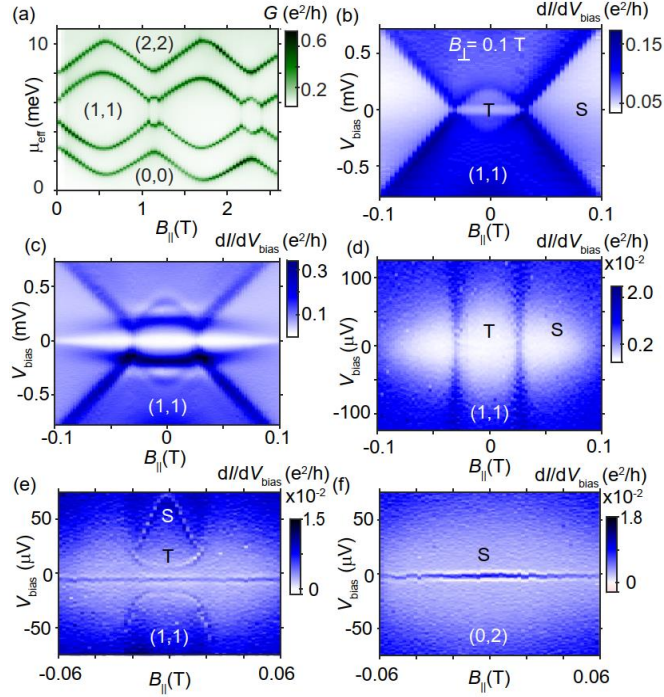

FIG. 4. (a) Conductance vs. DQD electrochemical energy and $B_\parallel$ obtained along the red gate vector in Fig. 2a crossing through the honeycomb triple points. $V_{bias}$ = 85 mV and $V_{BG}$ = -3.3 V. (b) Differential conductance vs. $V_{bias}$ and $B_\parallel$ inside the (1,1) honeycomb. A constant $B_\perp$ is applied to quench superconductivity. A triplet-singlet GS transition occurs at $B_\parallel$ = 25 mT due to the large orbital angular momentum of S(1,1) from the ring-like DQD symmetry. (c) Corresponding measurement with $B = 0$. (d) Corresponding supercurrent measurement where we note the absence of supercurrents. (e,f) Similar measurements in the (1,1) and (0,2) regimes of another orbital degeneracy. In panel e we note a finite supercurrent which does not decay with $B_\parallel$ in the same way as in panel f obtained where the GS is S(0,2). Also in panel e, the S and T labelling indicate excited (white) and ground (black) states in the corresponding normal regime as determined from cotunneling spectra (see Fig. S2).

*Transport via two-electron spin states in the superconducting gap.* Finally we investigate detuning of the DQD ring across the two-electron configurations (2,0)-(1,1)-(2,0)e as shown in Figs. 5(a,b). Here we find clearly resolved states inside the superconducting gap, where the latter extends beyond the measurement range [Figs. 5(c,d)]. These states show a direct correlation with the expected evolution of spin states in a DQD, although with a spin-triplet GS in (1,1)e due to the ring-like symmetry.

Results from a calculation of DQD orbital- and spin-state energies as a function of detuning near (2,0)-(1,1)e are shown in Figs. 5(e,f) (see the Supplemental Material [27] for details). The singlets S(1,1) and S(2,0) can interact via tunneling to form two hybridized singlets near the charge degeneracy point. At this point, the GS changes from singlet to triplet, where the states couple through spin-orbit interaction (SOI). The change in GS primarily involves T(1,1) and S(2,0) due to the exchange interaction. We note that at least one of the triplets cuts through a small avoided crossing, $\Delta_{ST} \approx 20$ µeV, of T(1,1)/S(2,0), which has been corrected for $R_S$ [27]. A reason for the very weak S-T interaction, despite a considerable spin orbit interaction, SOI $\approx$ 300 µeV [22], is that for T(1,1)-S(2,0) the interaction scales with the inter-dot tunnel coupling, $t$ [33–35]. Since the overlap integrals have different signs at the two connections in a DQD ring [24] the effective $t$ becomes strongly suppressed. From Fig. 5(b), we extract a hybridization gap, $\omega$ = 250 and 120 µeV from the interactions of S(1,1) with S(2,0) and S(0,2) respectively, suggesting an average $t \approx 60$ µeV ($t = \omega/2^{3/2}$). This value is indeed considerably smaller than the corresponding $t \approx 0.9$ meV obtained close to the point where the (1,1)e GS change into a singlet ($V_{BG}$ = -2.3 V) and where the ring-like symmetry is suppressed [Fig. 3(d)].

Figure 5(g) reveals regions with strong negative d$I$/d$V_{bias}$ at energies that seem to correlate with transport involving excited spin-states. Similar features are typically observed for subgap bound states in QD systems attached to two superconducting contacts [13], where the effect grows with the asymmetry of the source-drain coupling strengths [36]. Despite the small energies involved and the strong QD-lead couplings, we note that the states are very well resolved. The spectroscopic resolution is improved by studying co-tunneling transport inside the superconducting gap, which leads to a significantly reduced lifetime broadening.

Once the spin-triplet is GS, the supercurrent becomes notably weaker. A non-zero $B_\parallel$ rapidly shifts the lowest S(1,1) down in energy, which becomes GS in the measurement at $B_\parallel$ = 40 mT in Fig. 5(h). But, as pointed out earlier, no supercurrent is discernible for S(1,1), despite a correspondingly weak supercurrent involving S(2,0) and S(0,2). This finding indicates a suppression of non-local tunneling transport trough the DQD ring states, or possibly a mechanism related to orbital pair breaking resulting from the large orbital magnetic moment [37,38].

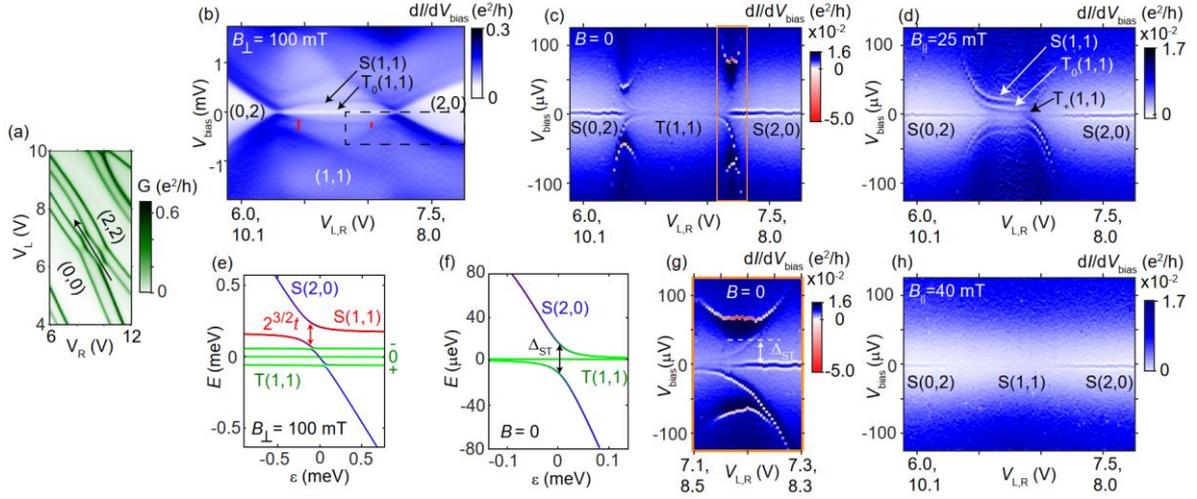

FIG. 5. (a) Transport measurements in this figure involve detuning of the ring-like DQD from (0,2) to (2,0) along the black arrow. (b) Transport in the normal state at $B_\perp = 100$ mT, with a $T_+(1,1)$ GS. The two red lines indicate avoided crossings of S(0,2)-S(1,1) and S(1,1)-S(2,0) from which we can extract the inter-dot tunnel coupling, $t$. (c) Corresponding supercurrent measurement with a suppressed supercurrent where T(1,1) is the GS. SOI results in an S/T avoided level crossing, where at least one state crosses the SOI-induced gap. (d) A $B_\parallel = 25$ mT spin splits T(1,1) states, and shifts the lowest S(1,1) down in energy. (e) Calculated energies of 2e states in a DQD with a single orbital in each QD and at $B_\perp = 100$ mT, corresponding to the dashed box in panel b. For simplicity, the energies are plotted as function of detuning of (1,1)/(2,0) states, where the detuning energy, $\varepsilon$, is set to 0 at the T/S crossing. (f) Calculation of the GS change from T(1,1) (left) to S(2,0) (right) at $B = 0$. (g) Corresponding GS change in the measurements obtained at the box indicated in panel c. (h) Measurement at $B_\parallel = 40$ mT with a spin-singlet GS where we note a supercurrent suppression in (1,1)e.

*Summary and conclusion.* Transport of Cooper pairs was studied in a coupled QD system with tunable symmetries. A suppression of supercurrent transport was found when the QDs couple in a way that minimizes hybridization and where the ground state is spin-triplet. The suppression was lifted when inducing a spin-singlet ground state by electrostatically changing the DQD interaction to strong hybridization. At orbital degeneracy, a broken particle-hole symmetry was found for odd electron occupations, where a supercurrent appeared in 1e but not in 3e. Transport studies near the S(2,0)-T(1,1) transition revealed clearly resolved sub-gap states that correlate with calculated two-electron spin states.

We envision that the electrostatically controlled spin ground state of this system could be used to study transport in unconventional superconductors, such as triplet or ferromagnetic superconductivity. Here, the DQD could directly probe the Cooper pair spin ordering. By inserting the DQD into one arm of a SQUID, it would also be possible to investigate predictions on the supercurrent-phase relationship in the case of interference in multiple channels, as well as the emergence of arbitrary phase (φ) junctions [39].

*Acknowledgements.* We thank Janine Splettstoesser, Athanasios Tsintzis, Rubén Seoane Souto, Fernando Dominguez, Andreas Baumgartner, and Martin Leijnse for helpful discussions. This work was carried out with financial support from NanoLund, the Swedish Research Council (VR), the Crafoord Foundation, and the Knut and Alice Wallenberg Foundation (KAW).

*claes.thelander@ftf.lth.se